\documentclass[twocolumn,aps,prd]{revtex4-2}
\usepackage{graphicx}
\usepackage{bm}
\usepackage{relsize}
\usepackage{amsmath,amsfonts,amsthm}
\begin{document}
\newcommand*{\bi}{\bibitem}
\newcommand*{\ea}{\textit{et al.}}
\newcommand*{\eg}{\textit{e.g.}}
\newcommand*{\zpc}[3]{Z.~Phys.~C \textbf{#1}, #2 (#3)}
\newcommand*{\plb}[3]{Phys.~Lett.~B \textbf{#1}, #2 (#3)}
\newcommand*{\phrc}[3]{Phys.~Rev.~C~\textbf{#1}, #2 (#3)}
\newcommand*{\phrd}[3]{Phys.~Rev.~D~\textbf{#1}, #2 (#3)}
\newcommand*{\phrl}[3]{Phys.~Rev.~Lett.~\textbf{#1}, #2 (#3)}
\newcommand*{\pr}[3]{Phys.~Rev.~\textbf{#1}, #2 (#3)}      
\newcommand*{\npa}[3]{Nucl.~Phys.~A \textbf{#1}, #2 (#3)}  
\newcommand*{\npb}[3]{Nucl.~Phys.~B \textbf{#1}, #2 (#3)}  
\newcommand*{\npbps}[3]{Nucl.~Phys.~B, Proc. Suppl. \textbf{#1}, #2 (#3)}  
\newcommand*{\ptp}[3]{Prog. Theor. Phys. \textbf{#1}, #2 (#3)}
\newcommand*{\ppnp}[3]{Prog. Part. Nucl. Phys. \textbf{#1}, #2 (#3)}
\newcommand*{\ptep}[3]{Prog. Theor. Exp. Phys. \textbf{#1}, #2 (#3)}
\newcommand*{\ibid}[3]{\textit{ ibid.} \textbf{#1}, #2 (#3)}
\newcommand*{\epjc}[3]{Eur. Phys. J. C \textbf{#1}, #2 (#3)}
\newcommand*{\jpg}[3]{J. Phys. G \textbf{#1}, #2 (#3)}
\newcommand*{\mpla}[3]{Modern Physics Letters A \textbf{#1}, #2 (#3)}
\newcommand*{\ra}{\rightarrow}
\newcommand*{\pippim}{\pi^+\pi^-}
\newcommand*{\kpkm}{K^+K^-}
\newcommand*{\kskl}{K^0_SK^0_L}
\newcommand*{\rf}[1]{(\ref{#1})}
\newcommand*{\be}{\begin{equation}}
\newcommand*{\ee}{\end{equation}}
\newcommand*{\bea}{\begin{eqnarray}}
\newcommand*{\eea}{\end{eqnarray}}
\newcommand*{\nl}{\nonumber \\}
\newcommand*{\rmd}{\mathrm d}
\newcommand*{\die}{e^+e^-}
\newcommand*{\jj}{\mathrm i}
\newcommand*{\ndf}{\mathrm{NDF}}
\newcommand*{\mev}{\mathrm{~MeV}}
\newcommand*{\cndf}{\chi^2/\mathrm{NDF}}
\newcommand*{\minuit}{\texttt{MINUIT}~}
\newcommand*{\w}{\sqrt s}
\newcommand*{\e}[1]{{\mathrm e}^{#1}}
\newcommand*{\ie}{\textrm{i.e.}}
\newcommand*{\dek}[1]{\times10^{#1}}
\newcommand*{\chdec}{K^+\to \pi^+\nu\bar\nu}
\newcommand*{\ndec}{K_L \to \pi^0\nu\bar\nu}
\newcommand*{\pdec}{K_L \to \pi^0\pio}
\newcommand*{\ppdec}{K_L \to \pi^0\piop}
\newcommand*{\dd}{\mathrm d}
\newcommand*{\pio}{\mathrm{A}_{2\pi}}
\newcommand*{\piop}{\mathrm{A}_{2\pi}^\prime}
\newcommand*{\gp}{{1{\mathrm s}}}
\newcommand*{\ep}{{2{\mathrm p}}}
\title{Pionium as a source of false events in the $\bm{K\to\pi\nu\bar\nu}$
decays}
\author{Peter Lichard}
\affiliation{
Institute of Physics
and Research Centre for Computational Physics and Data Processing, 
Silesian University in Opava, 746 01 Opava, Czech Republic\\
and\\
Institute of Experimental and Applied Physics,
Czech Technical University in Prague, 128 00 Prague, Czech Republic
}
\begin{abstract}
We suggest that the decay modes of kaons with a pion and a pionium 
($\pi^+\pi^-$ atom) in the final state can constitute a not yet considered 
background to the very rare decay $K\to\pi\nu\bar\nu$. In fact, a part of 
pioniums may escape the decay region before decaying into two $\pi^0$s (or 
to $\pi^0\pi^0\gamma$ in the case of excited pionium). To illustrate the
importance of this background, we show that it may even explain, under
some assumptions, the unexpected $K_L$ decay events that appeared 
in the KOTO experiment.
\end{abstract}
\maketitle
Two important experiments investigating the rare kaon decays in flight
are currently running. The main aim of both is to test the Standard Model and
to constrain new physics theories by precisely measuring the very rare kaon
decays into a pion and two neutrinos. The NA62 experiment at the CERN Super 
Proton Synchrotron \cite{NA62,NA62y2019a,NA62y2020a,NA62y2020b} deals with 
positively charged kaons $K^+$
and aims to collect enough $\chdec$ events to get a signal to background
ratio of 10:1. The Standard Model predicts the branching fraction 
\cite{terminology} for this decay \cite{theory}, 
\be
\label{chdec}
{\mathcal B}(\chdec)=(8.4\pm1.0)\dek{-11}. 
\ee 
The KOTO experiment \cite{koto} is being conducted at the Hadron Experimental 
Facility at the Japan Proton Accelerator Research Complex. 
It was designed to observe the decay $\ndec$ of long-lived neutral kaons. The 
theoretical branching fraction \cite{theory} is 
\be
\label{ndec}
{\mathcal B}(\ndec)=(3.4\pm0.6)\dek{-11}. 
\ee
Until recently, both rare kaon decay experiments proceeded as expected,
slowly pushing down the upper bounds of branching fractions. However, in 
September 2019, Satoshi Shinohara (on behalf of the KOTO Collaboration)
\cite{kotoanomaly} announced the presence of four events in the signal region
in the situation where mere 0.10$\pm$0.02 events were expected. Very soon,
several papers appeared, \eg, Refs. \cite{kitahara,bhupal,dutta}, 
aimed at finding new physics interpretations of this surprising result. The
analysis performed in Ref. \cite{kitahara} shows that if the four events
in the KOTO signal region were real, it would mean a branching fraction
of the underlying mechanism equal to
\be
\label{false}
{\mathcal B}(\ndec)_\mathrm{KOTO}=2.1^{+2.0(+4.1)}_{-1.1(-1.7)}\dek{-9} 
\ee
at the 68(95)\% confidence level.

The $\pi^+\pi^-$ atom, pionium (usually denoted as $\pio$), was discovered in 
1993 at the Institute of High Energy Physics at Serpukhov, 
Russia~\cite{pionium} and intensively studied in the Dimeson Relativistic 
Atomic Complex (DIRAC) experiment \cite{shortlife} at the CERN Proton 
Synchrotron. 
In these experiments, the pioniums were produced by the proton beam
impinging on a target. In the same target, the pioniums broke-up into
their constituents with approximately equal energies and small relative
momenta.

The decay of pionium to 
two neutral pions is dominant and the measured lifetime is \cite{shortlife}
\be
\label{shorttau}
\tau_{1\mathrm s}=3.15^{+0.28}_{-0.26}\dek{-15}~\mathrm s .
\ee 

The NA48/2 Collaboration at the CERN SPS \cite{NA48/2} studied 
$K^\pm\ra\pi^\pm\pi^0\pi^0$ decays and found an anomaly in the
$\pi^0\pi^0$ invariant mass distribution in the vicinity of 2$m_{\pi^+}$
that can be interpreted as the production of pioniums in the kaon decays and 
their subsequent two-$\pi^0$ decay. For our later considerations, it is 
important that the NA48/2 Collaboration in their seminal paper \cite{NA48/2}
determined the branching ratio
\be
\label{NA48ratio}
R=\frac{\Gamma(K^\pm\to \pi^\pm\pio)}{\Gamma(K^\pm\to\pi^\pm\pi^+\pi^-)}
=(1.61\pm0.66)\dek{-5}.
\ee

Another important discovery concerns the exited pionium. 
The DIRAC Collaboration observed \cite{longlife} so-called 
long-lived $\pippim$ atoms, which are the 2p atomic states ($\piop$)
with quantum numbers $J^{PC}=1^{--}$. 
Its lifetime was measured in Ref. \cite{longlife} with the result
\be
\label{longtau}  
\tau_{2\mathrm p}=0.45^{+1.08}_{-0.30}\dek{-11}~\mathrm s. 
\ee
Such a long lifetime is caused 
by the fact that the decay modes to the positive C-parity states $\pi^0\pi^0$ 
and $\gamma\gamma$ are now forbidden and the slow 2p$\to$1s transition
dominates. After reaching the 1s state, a decay to two $\pi^0$s quickly 
follows: $\pio^\prime\to\pio\gamma\to\pi^0\pi^0\gamma$.

The decay of the charged kaons into excited pionium $\pio^\prime$ has not
been reported yet. Neither has been the neutral kaon decay into any pionium.

In order to show that the pioniums may contribute to the background in the
$K\to\pi\nu\bar\nu$ experiments by 
important amount, we will estimate
it in the KOTO experiment. 
To this end, we need at least a crude estimate of the branching
fraction of decay $K_L\to\pi^0\pio$. We assume that the branching ratio
\[
\tilde{R}=\frac{\Gamma(K_L\to \pi^0\pio)}{\Gamma(K_L\to\pi^0\pi^+\pi^-)}
\]
has the same value as that for charged kaons \rf{NA48ratio}. Then we can
write the branching-fraction estimate
\bea
\label{bf}
{\mathcal B}(K_L\to \pi^0 \pio)&=&R\times{\mathcal B}(K_L\to \pi^0\,
\pi^+\pi^-)
\nl
&\approx& 2\dek{-6},
\eea
where we have also consulted Ref. \cite{pdg}. 

We will also consider possibility that the pionium that appears in the kaon
decays is not the ground state (1s) pionium, but its excited (2p) partner
and that the corresponding branching fraction are the same,
\be
\label{bfprime}
{\mathcal B}(K_L\to \pi^0 \piop)\approx 2\dek{-6}.
\ee
In what follows, we will pursue those two alternatives in parallel.

To simplify the reasoning, we will ignore the momentum spread of the $K_L$ 
beam in the KOTO experiment and  will use its peak value $P=1.4$~GeV/c 
\cite{koto}. The length of the KOTO signal region is $L$=1.7~m. Another quantity 
that enters the game is the pionium mass. It is given by $m_a=2m_{\pi^+}-b$, 
where $b$ is the binding energy. We will take its coulombic value, which can 
be calculated from the hydrogen-atom-like formula. One obtains $b=1.86$~keV
for the pionium ground state and $b=0.464$~keV for the excited 2p state.

The laboratory energy of pionium with mass $m_a$ is uniformly distributed in 
an interval, the bounds of which are given by the formula
\[
E_{a\pm}=\frac{1}{M}\left(EE_a^*\pm Pp_a^*\right),
\]
where $M$, $E$, and $P$ are the mass, energy, and momentum of the $K_L$,
respectively, $E_a^*=(M^2+m_a^2)/(2M)$, $p_a^*=(M^2-m_a^2)/(2M)$.
Numerically, $E_{a-}=0.497$~GeV and $E_{a+}=1.456$~GeV.

To simplify the consideration further \cite{mc}, we will assume that all
pioniums have the
same laboratory momentum, given as the mean value,
\bea
p_a&=&\frac{1}{E_{a+}-E_{a-}}\int_{E_{a-}}^{E_{a+}}\sqrt{E^2-m_a^2}\ \dd E
=\frac{1}{2(E_{a+}-E_{a-})}\nl
&\times&\left[E_{a+}p_{a+}-E_{a-}p_{a-}-m_a^2
\log\frac{E_{a+}+p_{a+}}{E_{a-}+p_{a-}}\right],
\eea
where $p_{a\pm}=\sqrt{E_{a\pm}^2-m_a^2}$. Numerically, $p_a=0.880$~GeV/c.

The probability that pionium travels the path $s$ without decaying is 
given by
\[
S(s)=\exp\{-s/l\},
\]
where $l$ is the mean decay length of pionium,
\be
\label{decayl}
l=\frac{p_a\tau}{m_a},
\ee
$\tau$ being the pionium mean lifetime. 
For the two types of pioniums, we get
$l_{1\mathrm s}=2.98^{+0.27}_{-0.25}~\mu$m and
$l_{2\mathrm p}=4.2^{+10.2}_{-2.8}$~mm. 

If we denote the length of the signal region as $L$, the mean survival 
probability of pionium at the point where it leaves the signal region is
\be
\label{survprob}
\bar S=\frac{1}{L}\int_0^LS(L-z)\dd
z=\frac{l}{L}\left[1-\exp\{-L/l\}\right].
\ee
Numerical values for two kinds of pionium are \cite{noteerrors}
\bea
\bar S_\gp&=&1.75^{+0.16}_{-0.15}\dek{-6},       \nl
\bar S_\ep&=&2.5^{+6.0}_{-1.7}\dek{-3}.          \nonumber
\eea

Multiplying these numbers by the assumed branching fractions \rf{bf} and
\rf{bfprime}, we obtain the branching fractions of $\pdec$ and $\pdec^\prime$ 
events that look like the $\ndec$ events because pioniums left the KOTO decay 
volume undecayed,
\bea
\label{bskoto}
{\mathcal B}_\gp&=& 3.50^{+0.32}_{-0.30}\dek{-12},\\
\label{bvkoto}
{\mathcal B}_\ep&=& 5.0^{+12.0}_{-3.4}\dek{-9}.
\eea
Branching fraction \rf{bskoto} is by 3 orders of magnitude smaller
than branching fraction \rf{false}, which characterizes the presence
of unexpected events in the KOTO signal region. Therefore, we will not
follow the 1s pionium option any longer.

A comparison of \rf{bvkoto} with \rf{false} suggests that the anomalous events
in the KOTO experiment could be explained as undecayed 2p pioniums. But to
make a realistic comparison, we must take into account that the experimental
efficiencies for the $\ndec$ and $\ppdec$ are different. The right quantity
that should be compared with \rf{false} is
\be
{\mathcal B}^\prime_\ep=R\times {\mathcal B}_\ep,
\ee
where $R$ is the ratio of efficiencies,
\be
\label{ratio}
R=\frac{\epsilon_{\pi^0\piop}}{\epsilon_{\pi^0\nu\bar\nu}}.
\ee
Recently, the authors of Ref. \cite{kitahara} considered a similar ratio,
\be
\label{ratiojap}
R(m_{X^0})=\frac{\epsilon_{\pi^0X^0}}{\epsilon_{\pi^0\nu\bar\nu}},
\ee
where $X_0$ is an invisible boson. They performed a Monte Carlo simulation
taking into account the experimental cuts, acceptances, and resolutions.
The dependence of ratio \rf{ratiojap} on the boson mass $m_{X^0}$ can be
obtained from the curve in the left pane of Fig. 2 \cite{kitahara}.
In principle, it should be possible to use their results and get 
ratio \rf{ratio} by setting $m_{X^0}$ to $m_{\piop}$. Unfortunately,
the value for $m_{\piop}$, which is close to 280~MeV, cannot be read 
from the curve. We will therefore make our own crude estimate of the
efficiency ratio \rf{ratio}.

The efficiencies are most influenced by the $p_T$ cut imposed in the KOTO
experiment. To suppress the events from the $K_L\to\pi^+\pi^-\pi^0$, a
$p_T$ momentum of $\pi^0$ greater than 130~MeV/$c$ is required over
the majority of the signal region; at the downstream edge, the cut is even 
higher (up to 150~MeV/$c$). This cut also suppresses the events with two-body
final states containing, besides the $\pi^0$, a massive particle $X^0$.
It may be an invisible light boson, as in Ref.~\cite{kitahara}, or 
a pionium. In these events, the maximum transverse momentum of $\pi^0$ is 
equal to the momentum of the outgoing decay products in the $K_L$ rest
frame,
\be
\label{ptmaxX}
p_{\mathrm{T,max}}=\frac{1}{2m_K}\sqrt{\lambda(m_K^2,m_\pi^2,m_{X^0}^2)}\ ,
\ee
where we use the usual notation
\[
\lambda(x,y,z)=x^2+y^2+z^2-2xy-2xz-2yz.
\]
The maximum transverse momentum of $\pi^0$ in the $\ndec$ events 
is higher because the lowest invariant mass of the two-neutrino system is zero
(neutrinos' masses neglected), namely,
\[
 p_{\mathrm{Tmax},\nu\nu}= \frac{m_K^2-m_\pi^2}{2m_K}=230.5~\textrm{MeV}/c.
\]

Given a $p_T$ cut, we can restrict the masses of $X^0$s that can be
detected as
\[
m_{\mathrm{X}^0}<\sqrt{m_K^2+m_\pi^2-2m_K\sqrt{m_\pi^2+p_{\mathrm{T,max}}^2}}\ .
\]
For $p_{\mathrm{T,max}}=130$~MeV/$c$, it implies $m_{\mathrm{X}^0}<281$~MeV.
It is obvious that $\epsilon_{\pi^0X^0}$ should go rapidly to
zero with $m_{X^0}$ approaching 281~MeV. Figure 2 in Ref.~\cite{kitahara}
confirms that.

In order to get estimates of the efficiencies, we first randomly generate 
the particle momenta within the $\pi^0\nu\bar\nu$ system in the
$K_L$ rest frame using the \texttt{GENBOD} program from the old CERN 
Program Library \cite{genbod}. Counting the events with the $\pi^0$
transverse momentum greater than the KOTO cut of 130~MeV/$c$, we get an
efficiency $\epsilon_{\pi^0\nu\bar\nu}$ of 22\%.

Next, we consider the 2p pionium with the Coulombic binding energy, the mass
of which is $m_a=279.14$~MeV. Using the formula \rf{ptmaxX} we get 
$p_{{\mathrm{Tmax},\pi^0}}$=132.0 MeV/$c$. Assuming isotropic decay in the
$K_L$ rest frame, we can estimate the portion of events with $p_{T,\pi^0}$
greater than 130~MeV/$c$ as
$\epsilon_{\pi^0\textrm{A}^\prime}=(132-130)/132=1.5\%$. It means
an efficiency ratio \rf{ratio} of
\[
R(\textrm{Coul.})=0.069
\]
and an efficiency corrected branching ratio
\[
{\mathcal B}^\prime_\ep(\textrm{Coul.})=0.34^{+0.83}_{-0.23}\dek{-9}.
\]
A comparison with \rf{false} shows that if the experimental cut on transverse 
momentum of $\pi^0$ is taken into account, the 2p pioniums with the Coulombic 
binding energy cannot be a source of the unexpected events in the KOTO 
experiment. 

Yet, no direct measurement of the 2p pionium mass or binding energy exists.
From its decay chain, we know that its mass must be greater than $2m_{\pi^0}$.
The binding energy $b$ should thus be smaller than $2(m_{\pi^+}-m_{\pi^0})
\approx9.19$~MeV. It is possible 
\cite{manifestation} that the binding energy of $\piop$ may be around
9~MeV \cite{noteuretsky}. 
In that case, we have $m_a=270.14$~MeV, $p_{\mathrm{Tmax,A^\prime}}=
139.0$~MeV/$c$, $\epsilon_{\pi^0\textrm{A}^\prime}=(139-130)/139=6.5\%$,
and
\[
R(b=9~\textrm{MeV})=0.29.
\]
The efficiency corrected branching fraction now comes out as
\[
{\mathcal B}^\prime_\ep(b=9~{\textrm MeV})=1.5^{+3.5}_{-1.0}\dek{-9},
\]
which is, within the errors, compatible with the branching fraction
\rf{false} characterizing the unexpected KOTO events.

The KOTO Collaboration continues in analyzing its older data
and taking new ones \cite{shimizu}. 
One of the four events reported in \cite{kotoanomaly} has been shown to
be due to a mistake, so only three good candidate events remain. The new
background estimate is $1.05\pm0.28$ events. The number of expected
Standard Model events is 0.04. 

It is interesting that the KOTO experiment sees more $K^\pm$ than expected. 
It may be worth remaining that the $K_L\to K^\pm e^\mp \bar\nu(\nu)$ decays 
\cite{betak} have not been observed yet. 
Nevertheless, the KOTO Collaboration considered them \cite{taku} as a source
of background and varied the branching fraction in reasonable bounds. A
negligible contribution was found.
 
To conclude, we have suggested that events in which the kaon decays into
pionium, which then leaves the decay region without decaying, may contribute
to the background in the $K\to\pi\nu\bar\nu$ experiments. In our opinion,
this source of background  should be taken into account when analyzing
the existing experiments or planning new ones \cite{klever}. However, to get
reliable estimates of this background, the discovery and measurement
of the following decays will be very valuable: $\pdec$ (a neutral 
kaon partner of the charged kaon decay already observed \cite{NA48/2}),
$\ppdec$, and $K^\pm\to\pi^\pm\piop$. 

As an illustration of the idea, we have shown that, under some assumptions, 
the production of 2p pioniums $\piop$ in the $K_L$ decays and their subsequent 
escaping from the signal region may explain the production of unexpected events
reported by the KOTO Collaboration \cite{kotoanomaly}. Our key assumptions,
which may or may not be correct, have been as follows: (i) main contribution 
comes from pioniums that are in the 2p state, which lives much longer than 
the ground state; (ii) the branching 
ratio of $K_L\to\pi^0 \piop$ to $K_L \to \pi^0 \pi^+ \pi^-$ is the same as 
that of $K^+ \to \pi^+ \pio$ to $K^+ \to \pi^+\pi^+ \pi^-$, which was
determined by the NA48/2 Collaboration \cite{NA48/2};
(iii) the $\piop$ binding energy is larger than the Coulombic one, which 
helps pioniums to overcome the experimental $p_T$ cut.

\begin{acknowledgments}
I thank Teppei Kitahara for pointing me toward the importance of efficiencies
and Augusto Ceccucci, Matthew Moulson, and Taku Yamanaka for useful 
correspondence.

This work was partly supported by the Ministry of Education, Youth and
Sports of the Czech Republic Inter-Excellence Project No. LTI17018 and 
European Regional Development Fund Project No. 
CZ.$02.1.01/0.0/0.0/16\_019/0000766$.
\end{acknowledgments}

\end{document}